\documentclass[twocolumn,aps,pra,amsmath,amssymb,amsfonts,superscriptaddress,notitlepage,reprint,longbibliography,floatfix%,linenumbers
]{revtex4-1}
\usepackage{graphicx}
\usepackage{dcolumn}
\usepackage{bm}
\usepackage{amssymb}
\usepackage[svgnames]{xcolor}
\usepackage[colorlinks=true,
            linkcolor=red,
            urlcolor=blue,
            citecolor=DarkGreen]{hyperref}
\usepackage[english]{babel}

\begin{document}

\title{Parity Detection of Propagating Microwave Fields}
\author{Jean-Claude~Besse}\email{jbesse@phys.ethz.ch}
\author{Simone~Gasparinetti}\thanks{Current address: Quantum Technology Laboratory, Chalmers University of Technology, SE-412 96 Göteborg, Sweden}
\author{Michele~C.~Collodo}
\author{Theo~Walter}
\author{Ants~Remm}
\author{Jonas~Krause}
\author{Christopher~Eichler}
\author{Andreas~Wallraff}\affiliation{Department of Physics, ETH Zurich, CH-8093 Zurich, Switzerland}

\date{\today}

\begin{abstract}
The parity of the number of elementary excitations present in a quantum system provides important insights into its physical properties. Parity measurements are used, for example, to tomographically reconstruct quantum states or to determine if a decay of an excitation has occurred, information which can be used for quantum error correction in  computation or communication protocols.
Here we demonstrate a versatile parity detector for propagating microwaves, which  distinguishes between radiation fields containing an even or odd number $n$ of photons, both in a single-shot measurement and without perturbing the parity of the detected field. We showcase applications of the detector for direct Wigner tomography of propagating microwaves and heralded generation of Schr{\"o}dinger cat states. This parity detection scheme is applicable over a broad frequency range and may prove useful, for example, for heralded or fault-tolerant quantum communication protocols.
\end{abstract}

\maketitle

\section{Introduction}

In quantum physics, the parity $P$ of a wavefunction $\psi$ governs whether a system has an even or odd number of excitations $n$. The parity affects, for example, the system's statistical properties such as the likelihood of transitions occurring between distinct quantum states~\cite{Binney2013,Atkins2017}. An ideal measurement of the parity $P$ of a system distinguishes states with even ($0,2,4,...$) from states with odd $n$ ($1,3,5,...$), while not providing any other information about the precise value of $n$.

In superconducting circuits, for example, the parity of the number of photons stored in a microwave cavity is determined either by direct measurements~\cite{Sun2014}, providing immediate access to the value of $P$, or indirect measurements \cite{Hofheinz2009}, requiring the reconstruction of $P$ from another measured quantity. Direct measurements of the parity are frequently used to reconstruct quantum states of radiation fields stored in microwave cavities \cite{Haroche2007,Sun2014,Ofek2016}. However, parity measurements of propagating quantum radiations fields, which can be used as the carriers of information in quantum networks, have just recently been realized in the optical domain~\cite{Hacker2019} with neutral atom based systems~\cite{Ritter2012}, while experimental realizations in the microwave domain are still lacking. Multi-photon quantum non-demolition (QND) measurements of itinerant microwave fields are an essential element for error detection~\cite{Bergmann2016} and error correction in information processing networks as they provide a path towards detecting photon loss.

Parity measurements also play an important role in protocols for error correction in quantum information processing \cite{Terhal2015n,Gottesman2001a} and quantum communication applications \cite{Kimble2008}. In that context parity measurements have been demonstrated, for example, with superconducting qubits for measurement based entanglement generation \cite{Riste2013}, for elements of error correction \cite{Riste2015,Kelly2015,Takita2016}, and entanglement stabilization \cite{Andersen2019}, an experiment which was also performed with ions \cite{Linke2017a,Negnevitsky2018}.

\section{Parity detection scheme}
The parity detector for propagating microwave fields introduced here is based on a cavity QED system realized in superconducting circuits. We characterize the detector performance by measuring the parity of single and multi-photon states distributed sequentially over multiple time bins as generated by a true single microwave photon source. We illustrate the use of the detector to directly evaluate the Wigner function of propagating fields of single photons and their coherent superpositions with vacuum by measuring their displaced parity. Finally, we highlight the single-shot and QND nature of the parity detector by heralding propagating, microwave-frequency Schr\"odinger cat states, with a definite even or odd parity, from incident coherent states with varying amplitude $\left|\alpha\right|$.

To measure the parity $P$ of a propagating microwave field, we engineer a controlled phase gate between a superconducting transmon qubit embedded in a cavity, acting as an ancilla, and itinerant microwave photons~\cite{Duan2004,Besse2017,Kono2018} acting as the control field. We realize this gate by tuning the first, $|e\rangle$, to second excited-state, $|f\rangle$, transition of the transmon qubit, $\omega_\mathrm{ef}/(2\pi)=5.9\,$GHz, into resonance with the fundamental mode of a cavity. The ground, $|g\rangle$, to first excited state transition $\omega_\mathrm{ge}$ is detuned by the anharmonicity $\alpha/(2\pi)=\left(\omega_\mathrm{ef}-\omega_\mathrm{ge}\right)/(2\pi)=-220\,$MHz, from the cavity mode. Thus, a vacuum Rabi mode splitting, of size $2g_1/(2\pi)=76\,$MHz, occurs if and only if the transmon is prepared in the excited state $|e\rangle$ \cite{Besse2017}. This ancilla-based scheme allows for the quantum-non-demolition measurement of the photon-number parity of the propagating field reflected off the input of the detector, a feature which we demonstrate explicitly here.

\begin{figure}[!tb]
\includegraphics{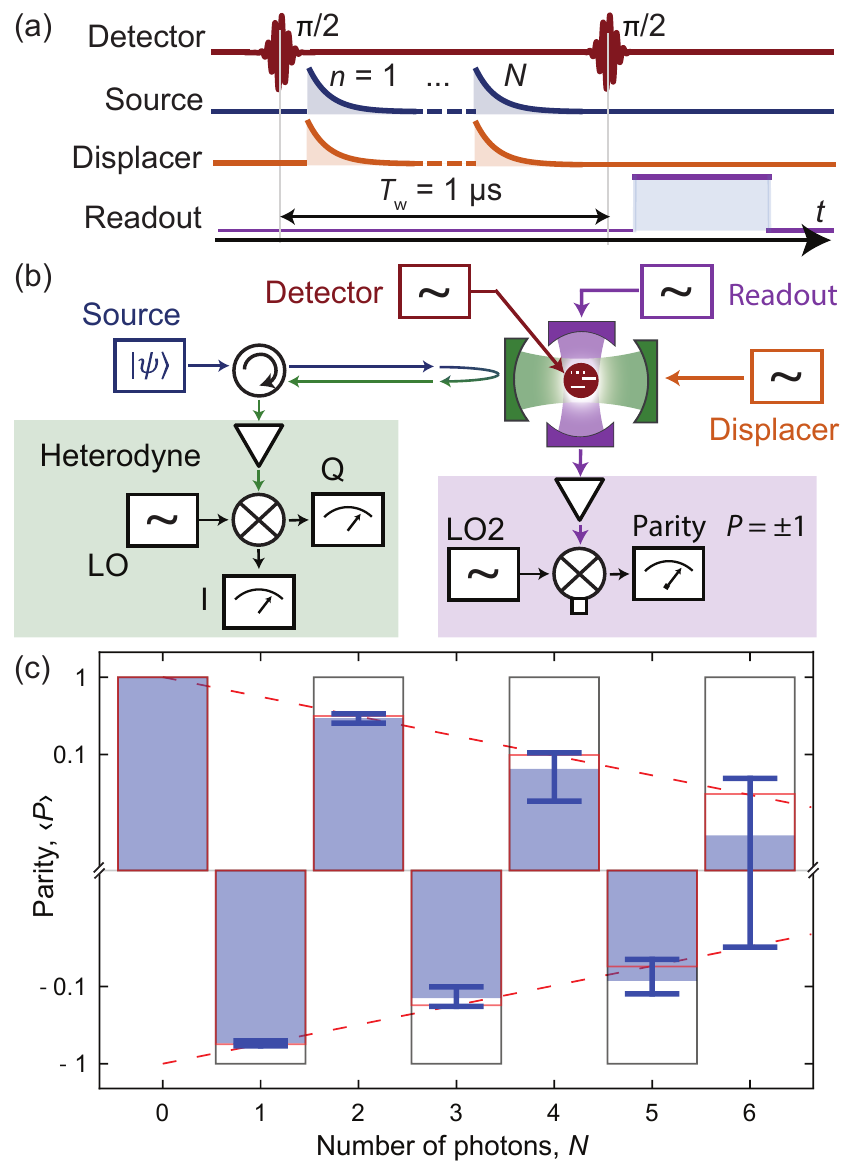}
\caption{\label{fig:1}{\bfseries Experimental setup and parity measurements.}
(a) Parity detection pulse sequence: Ramsey pulses applied to detector qubit (red), pulse train of $N$ spontaneously emitted photons (blue), externally applied coherent mode-matched displacer field (orange), and readout pulse (purple).
(b) Radiation coming from a source is reflected off of a cavity (green) coupled to a transmon (red) acting as the detector. The source is either a single photon emitter or a pulsed microwave generator with amplitude and phase control. Dispersive readout of the transmon, assisted by an additional cavity (purple), yields the photon parity (purple box). Fields inside the detector cavity can be displaced in-situ by applying an additional coherent tone (orange). Standard heterodyne detection of the $(I,Q)$-quadratures of the reflected light field (green box) is performed with a local oscillator (LO).
(c) Measured parity $\langle P \rangle$ (blue bars, on positive and negative log scales) for a train of $N$ single photon pulses. Ideal value of $\langle P \rangle$ (dark gray wireframes) and $\langle P \rangle$ considering finite transmission efficiency $\eta=78\,\%$ between source and detector (red wireframes and dashed lines). Error bars indicate the statistical standard deviation of $\pm 4\%$ of the parity.
}
\end{figure}

We arm the parity detector for a time $T_w=1\,\mu$s, shorter than both the lifetime $T_1 = 4.5\,\mu$s and dephasing time $T_2^\star = 3.5\,\mu$s of the detector transmon qubit, by defining a Ramsey sequence formed by two $\pi/2$ pulses separated by $T_w$, Fig.~\ref{fig:1}(a). Each photon impinging on the detector input during the time $T_w$ imparts a phase shift of $\phi=\pi$ on the superposition state $(|g\rangle + e^{i\phi}|e\rangle)/\sqrt{2}$ of the transmon qubit created by the first $\pi/2$ pulse~\cite{Besse2017}. As $e^{i \phi}$ is $2\pi$-periodic, the Ramsey sequence encodes the parity of the total number of scattered photons in the qubit population after the second $\pi/2$ pulse, leaving the transmon in $|e\rangle$ for even $\langle P \rangle=+1$ or in $|g\rangle$ for odd $\langle P \rangle=-1$. A schematic of the measurement setup is shown and discussed in Fig.~\ref{fig:1}(b), the sample and the wiring are presented in Appendix~\ref{app:Sample}.

\section{Parity measurements}
We examine the performance of this parity detector using a well characterized spontaneous-emission, single-photon source~\cite{Pechal2016}, operated on a separate chip. This source is capable of creating phase coherent superpositions of vacuum, $|0\rangle$,  and single photon, $|1\rangle$, states in a single time bin with a pulse bandwidth $\kappa_\mathrm{p}/(2\pi)=2\,$MHz, which is small compared to the effective parity detector bandwidth, set by the linewidth $\kappa_\mathrm{eff}/(2\pi)=30\,$MHz of the detector cavity. In this way the phase imparted on the detector qubit by each photon is well defined. Since the photon pulse length, $1/\kappa_\mathrm{p}$, is short compared to $T_w$, it is fully detected by the Ramsey sequence during which the detector is armed.

We operate the single photon source to emit sequences of $N=0,1,...,6$ pulses each containing a single photon Fock state $|1\rangle$~\cite{Peng2016d,Pechal2016} travelling towards the detector and record the average parity $\langle P \rangle$ of the pulse train as indicated by our detector. We take the finite phase coherence time $T_2^\star$ of the qubit and its readout fidelity into account to linearly map the measured qubit excited population $P_e$ to a parity value $\langle P \rangle$ using reference traces consisting of Ramsey sequences with the same $T_w$, Appendix~\ref{app:pop-parity}.

We observe the measured parity $\langle P \rangle$ (blue bars) changing sign, as expected, for each added single photon pulse establishing the detector's capability to discriminate even ($\langle P \rangle=+1$) from odd ($\langle P \rangle=-1$) photon number parity, see  Fig.~\ref{fig:1}(c) with the ideal result indicated by dark gray wireframes. The contrast in the measured $\langle P \rangle$, plotted on a logarithmic scale, reduces in good agreement with $(1-2\eta)^N$ (dashed red line) due to the finite transmission efficiency $\eta = 78\,\%$ between the source and the detector, Appendix~\ref{app:pop-parity}.
We note that these losses are external to the parity detector and independent of the detection event.

\section{Wigner tomography}
Measuring the expectation value of the parity operator $P$ of a radiation field described by a wave function $|\psi\rangle$, or a corresponding density matrix $\rho=|\psi\rangle\langle \psi|$, displaced by an operator $D_\alpha = \mathrm{exp}(\alpha^\dagger a - \alpha a^\dagger)$
directly yields the value of the Wigner function
\begin{equation}\label{eq:Wigner}
\frac{\pi}{2} W(\alpha) = \langle \psi | D^\dagger_\alpha P D_\alpha | \psi \rangle =  \mathrm{Tr} (P D_\alpha \rho D^\dagger_\alpha )
\end{equation}
at the point $\alpha$ in phase space~\cite{Lvovsky2002,Hofheinz2009,Shalibo2013}. Here $a$ is the photon annihilation operator. With our parity detector, we directly measure $W(\alpha)$, realizing the displacement operation of the field to be detected by applying a mode-matched coherent field to a second weakly coupled input of the detector cavity-qubit system, [Fig.~\ref{fig:1}(a,b)].

\begin{figure}[b]\setlength{\hfuzz}{1.1\columnwidth}
\begin{minipage}{\textwidth}
\includegraphics{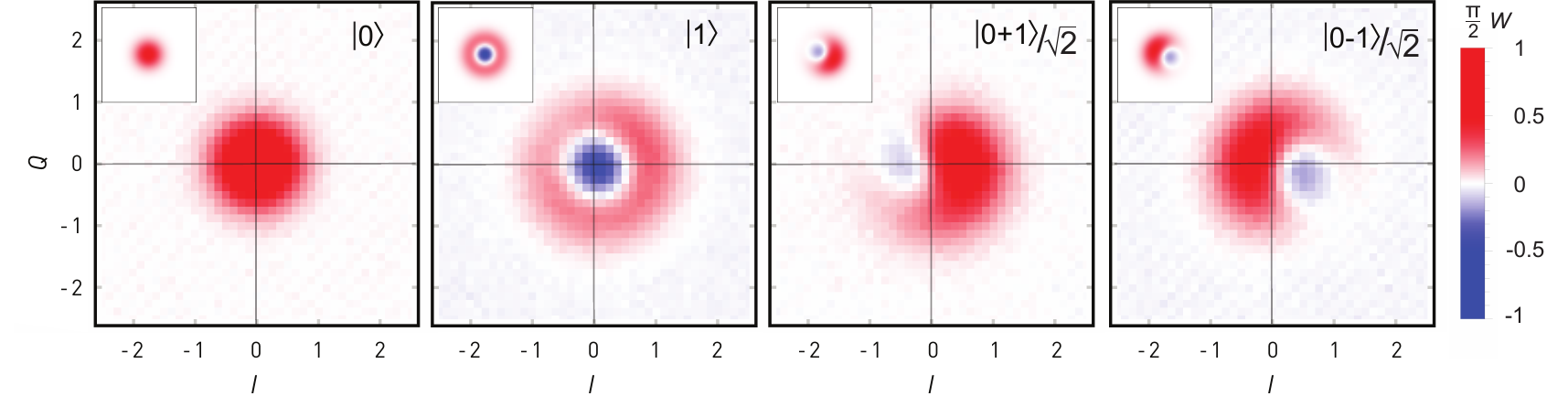}
\caption{\label{fig:3}{\bfseries Wigner tomography by displaced parity measurement.} Wigner tomography $\frac{\pi}{2} W(I+\mathrm{i}Q)$ of the vacuum state $|0\rangle$, the single photon Fock state $|1\rangle$, and their coherent superposition $|0{+}1\rangle/\sqrt{2}$ and $|0{-}1\rangle/\sqrt{2}$. Insets show the expected results for the ideal states taking inefficiencies into account (see text).
}
\end{minipage}
\end{figure}

We illustrate the method on the vacuum $|0\rangle$ and single photon Fock states $|1\rangle$ as well as their phase-coherent superpositions $(|0\rangle \pm |1\rangle)/\sqrt{2}$ created by our single photon source. All measurements of $\frac{\pi}{2} W(\alpha)$ were evaluated for a 41x41 points grid of the in-phase $I$ and out-of-phase $Q$ quadratures of the mode-matched displacer pulse defining the amplitude $\alpha = I+\mathrm{i}Q$.
We find excellent agreement between the measured Wigner functions and expected ones, for the created states, Fig.~\ref{fig:3}.
We also observe characteristic negative regions, a feature of quantum signals, in the measured Wigner functions. In particular, we record the pure single photon state $|1\rangle$ as a radially symmetric Wigner function with value $\frac{\pi}{2} W(0,0) = -0.55$ at the origin of the phase space. For $|0\pm 1\rangle/\sqrt{2}$, we note that the path difference between source and displacer lines causes a phase rotation of approximately $-25\,$degrees relative to the $I$-axis, which we chose not to correct for in the data analysis.

We reconstruct the most-likely density matrix for each of these states, imposing physicality with positive semi-definite programming as well as taking the experimentally determined finite mode-matching fidelity $F_\text{mm}=84\%$ and transmission efficiency $\eta=78\%$ into account and find an average fidelity of 95\% with respect to the ideal states, see Appendix~\ref{app:wigner-model}.

We note that with our detection method we are also able to determine the joint parity of radiation fields occupying distinct time bins within the same detection time window $T_w$ and thus perform joint Wigner tomography on those fields. Since our current photon source did not allow us to create states with entanglement shared across different time bins, we defer this discussion to later work.

\section{Heralding of cat states by parity detection}
Finally, we illustrate the quantum-non-demolition and single-shot character of this well characterized parity detection scheme by projecting an incident itinerant coherent state $|\alpha\rangle$, having a Poisson distributed photon number, into an eigenstate of the parity operator $P$ by parity detection. We experimentally demonstrate that this process creates heralded propagating even/odd parity cat states when conditioned on the single-shot parity measurement outcome.

We characterize the quantum properties of the reflected field after its interaction with the parity detector by measuring the statistical moments $\langle a^{\dagger n} a^m \rangle$~\cite{Eichler2011,Eichler2012,Eichler2012b} up to order $n,m\leq 7$ (see Appendix~\ref{app:cats-moments}). In principle, we could have used a second parity detector for that purpose. By comparing the noise in the detection chain with near-quantum-limited linear amplification to the signal level of a single photon, we extract an overall quantum efficiency of the phase-preserving heterodyne detection of $\eta_\text{het} = (1+N_0)^{-1} = 23\%$, with $N_0=3.3$ photons of added noise~\cite{Eichler2012}.

\begin{figure}
\includegraphics{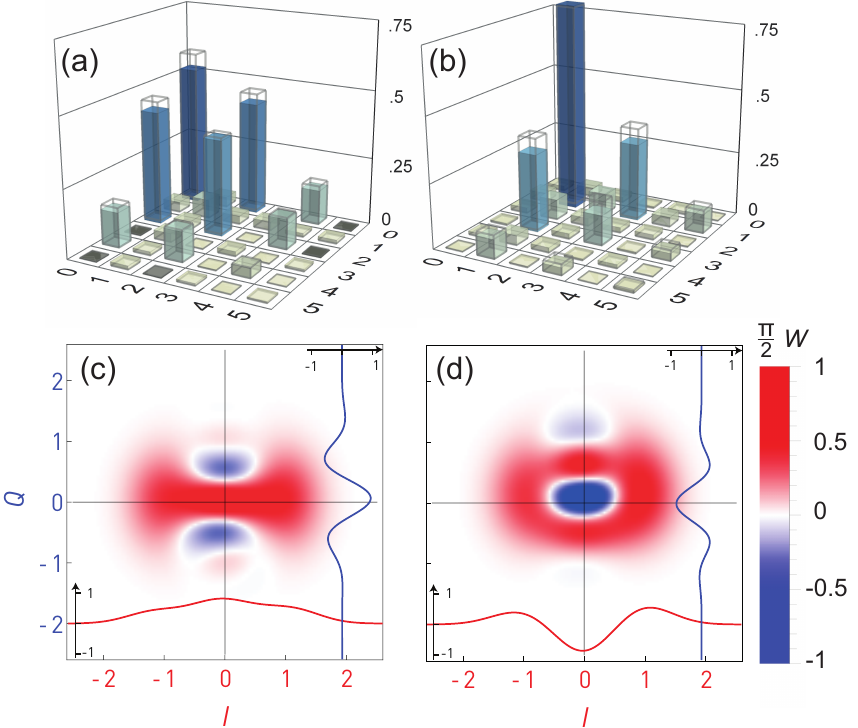}
\caption{\label{fig:4}{\bfseries Propagating microwave cat states.} (a,b) Real part of density matrices $\mathrm{Re}(\rho_{ij})$ for (a) the even cat state, (b) the odd cat state with amplitude $\alpha = 1.06$. Ideal cat states are shown as wireframes. Wigner functions reconstructed from the measured moments of the (c) even and (d) odd cat state for $\alpha = 1.06$. Cuts along the $I$-axis [$\frac{\pi}{2}W(I,Q=0)$, red solid line] and $Q$-axis [$\frac{\pi}{2}W(I=0,Q)$, blue solid line].
}
\end{figure}

{
In each single measurement of the quadratures of the field reflected from the detector, we also register the weight-integrated quadrature of the single-shot readout~\cite{Walter2017} of the transmon qubit embedded in the cavity of the parity detector in 120$\,$ns with a fidelity of $F_\text{ro}=94$\%. The single-shot correlations between the post-detection field and detector state, allow us to reconstruct the most likely state $|s\rangle$ of the full system after the interaction of the input field with the detector. From the data, we evaluate the density matrices $\rho$ of the radiation field projected onto the even (odd) detector parity subspace corresponding to the detector qubit states $|e\rangle$ ($|g\rangle$). Performing this projection, we correct for the finite qubit readout infidelity of $1-F_\text{ro}=6$\%~\cite{Eichler2012b}. In the analysis we truncate the Hilbert space of the radiation field at $n=5$. For the input amplitude $\alpha=1.06$, for example, we observe that the $+1)$ ($-1$) parity state of the detector heralds the radiation field in an even (odd) cat state, containing entries predominantly in the even
\parfillskip=0pt\par}
\newpage
\noindent
(odd) rows and columns of the corresponding density matrices, Fig.~\ref{fig:4}(a,b). This is the expected consequence of the cat states $|\mathrm{cat}_\mathrm{e,o}\rangle = (|\alpha\rangle \pm |{-}\alpha\rangle)/\mathcal{N}_{\alpha}^{(\pm)}$ being eigenstates of the parity operator $P$ with eigenvalue $\pm 1$. $\mathcal{N}_{\alpha}^{(\pm)} = [2(1\pm e^{-2|\alpha|^2})]^{1/2}$ is a normalizing factor.

The fidelity with respect to the ideal even (odd) post-detection cat states for this value of $\alpha$ (wireframes) is $F_\mathrm{e(o)}=0.88 (0.93)$, when correcting for the finite readout infidelity of 6\%. Without correcting for the readout infidelity, the fidelity is $\tilde{F}_\mathrm{e(o)}=0.86 (0.91)$. $F_\mathrm{e(o)}$ is limited by the coherence of the qubit, as well as by its steady-state thermal population $P_e^\mathrm{th}=4\%$. The latter could be reduced substantially by performing qubit reset~\cite{Magnard2018}.

The Wigner functions after parity detection, $W(\alpha)$, which we calculate from the most likely density matrices, show the expected features: two positive regions centered at $\pm \alpha$ along the $I$-axis (red line), and fringes along the $Q$-axis (blue line), with values of opposite sign at the origin, Fig.~\ref{fig:4}(c,d). The reconstructed density matrices show a high contrast between the parity of the even and odd cat state, which in other works is referred to as the fringe visibility~\cite{Hacker2019}, which we evaluate to $V=P_\mathrm{even}-P_\mathrm{odd}=1.7$ (1.6 without correction for the readout error) for our data, close to the ideal value of 2. The high-fidelity projection of the detected state into an eigenstate of the parity operator demonstrates the quantum non-demolition nature of the presented parity detection scheme. Similar measurement-induced collapse of coherent states into cat states has been reported recently with propagating optical fields at significantly lower parity contrast~\cite{Hacker2019} and for stationary microwave fields trapped in cavities~\cite{Vlastakis2013}.

\begin{figure}
\includegraphics{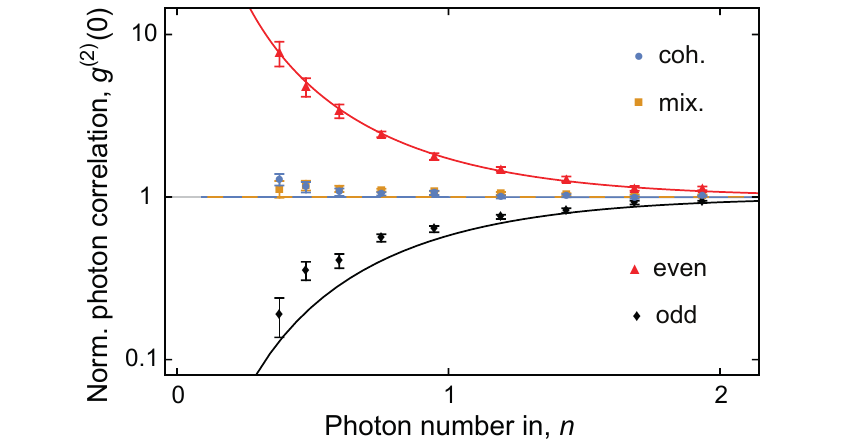}
\caption{\label{fig:5}{\bfseries Power dependence of photon correlations.} Normalized zero-time photon correlation $g^{(2)}(0)$ on a log scale vs average number of photons $n$ in the coherent tone applied to the input of the parity detector, for the coherent state $|\alpha\rangle$ (blue dots), a statistical mixture of $|\alpha\rangle$ and $|{-}\alpha\rangle$ (yellow dots), the even cat state (red triangles) and the odd cat state (black diamonds). Calculated values for the ideal states are shown as solid and dashed lines. Error bars indicate the statistical standard deviation of the data.
}
\end{figure}

We further explore the nonclassical properties of the generated cat states for mean photon numbers of up to $n = 2$ by extracting the normalized zero-time second order correlation function $g^{(2)}(0) = \langle {a^\dagger}^2a^2\rangle/\langle a^{\dagger}a\rangle^2$. $g^{(2)}(0)$ is directly computed from the measured moments, in contrast to the data plotted in Fig.~\ref{fig:4} for which maximum likelihood estimation was used.
We present $g^{(2)}(0)$ for four characteristic cases in Fig.~\ref{fig:5}: not operating the detector (blue), operating the detector in single-shot mode to create even (red) and odd cat states (black), and not distinguishing the parity detector states and thus creating a statistical mixture of cat states (orange symbols).
We observe a continuous transition: in the low power regime the odd cat state is anti-bunched, $g^{(2)}(0) \sim 0.2$, as it is composed predominantly of the single photon Fock state~\cite{Daiss2019}, and the even cat state is strongly bunched, $g^{(2)}(0) \sim 8$, as it consists mostly of vacuum and 2-photon components. Both the coherent state and the mixture obey poissonian statistics with $g^{(2)}(0) \sim 1$. For $\left|\alpha\right|^2>1$, the measured $g^{(2)}(0)\rightarrow 1$ for all states. This convergence with increasing power towards the value of the statistical mixture is expected for all moments of the cat states. Distinguishing cat states from classically mixed states thus becomes more difficult with increasing amplitude using the heterodyne detection technique. While this limits the maximum $\left|\alpha\right|$ characterized in this set of experiments, we expect our parity detector to work up to larger photon numbers. Experimentally, we have performed Wigner tomography of coherent states with up to $\left|\alpha\right|^2 = 10$ photons in the cavity, showing no degradation of performance. This is consistent with a parity detection process, though not sufficient to prove that the measurement operator is the expected one. The photon number at which non-linear effects become appreciable in the detection process remains to be determined. However, since keeping $\left|\alpha\right|^2 < 8$ ensures that the average population in the cavity is less than 1 at all times, we expect to be able to generate cat state of at least this order of magnitude.

\section{Conclusion}
We realized a parity detector for multi-photon itinerant microwave fields in the quantum regime.
We illustrated the use of the parity detector for direct Wigner tomography of propagating quantum microwave radiation fields at the single photon level. Single-shot projection onto parity eigenstates allowed us to demonstrate the non-demolition nature of the detection by generating heralded microwave-frequency cat states from coherent input fields.

QND parity detection of itinerant microwaves could facilitate connecting nodes of a quantum network faithfully in the presence of finite loss in the channel. Encoding quantum information in states with a given parity and performing joint parity measurements of time-bin encoded fields efficiently in single-shot, allow the parity detector to signal photon loss, i.e. taking the radiation out of the parity subspace, without providing information in the basis of encoding. Using the detector presented here may allow for repeated parity measurements and stabilization of a given parity subspace~\cite{Sun2014,Kelly2015,Liu2016c,Andersen2019}.

\section{Acknowledgments}
This work was supported by the European Research Council (ERC) through the “Superconducting Quantum Networks” (SuperQuNet) project, by the National Centre of Competence in Research “Quantum Science and Technology” (NCCR QSIT), a research instrument of the Swiss National Science Foundation (SNSF), and by ETH Zurich.

\appendix

\section{Sample fabrication and experimental setup}\label{app:Sample}
\begin{figure}[!b]
\includegraphics[width=\columnwidth]{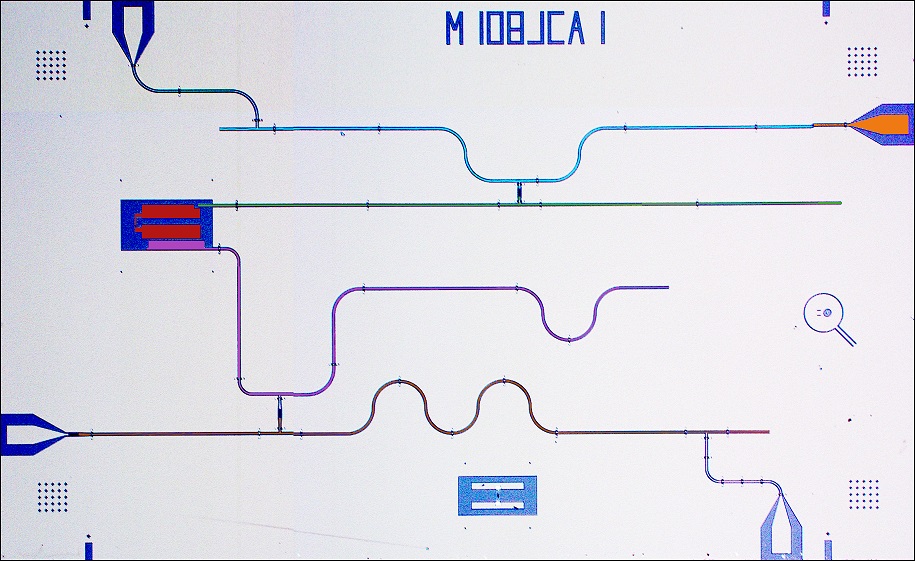}
\caption{\label{fig:sample}{\bfseries False color micrograph of the sample.} A transmon qubit (red) is coupled to the detection cavity (green) and its Purcell filter (cyan), as well as to a readout cavity (purple) and a corresponding Purcell filter (brown). A weakly coupled input port (orange) allows for displacing the field in the detection cavity. Silicon is shown in dark and Niobium in light gray.
}
\end{figure}

\begin{figure}
\includegraphics[width=\columnwidth]{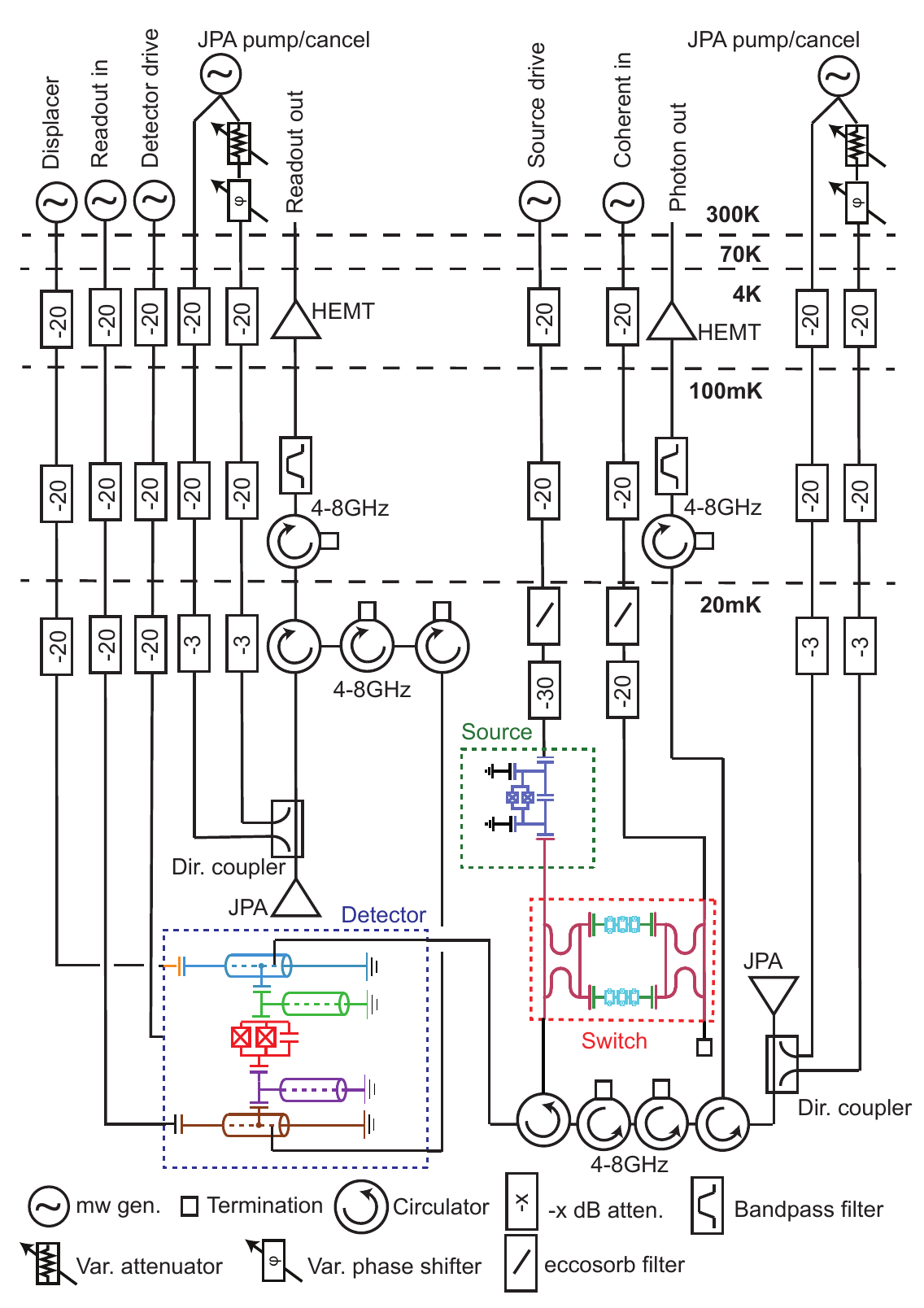}
\caption{\label{fig:setup}{\bfseries Schematic of experimental setup.} All microwave lines are inner/outer DC-blocked at the room temperature flange of the cryogenic system. Source and switch are physically on the same sample~\citep{Pechal2016}. DC cabling for applying external magnetic fields with one coil each on the switch sample holder and on the detector are not shown.
}
\end{figure}

The sample, shown in Fig.~\ref{fig:sample}, is fabricated on a 4.3$\,$mm x 7$\,$mm silicon chip. All elements except for the Josephson junctions are patterned in a 150$\,$nm-thick sputtered niobium film using photolithography and reactive ion etching. The Josephson junctions are fabricated in a separate step using electron-beam lithography and shadow-evaporation of aluminum in an electron-beam evaporator. As sketched in Fig.~\ref{fig:1}(b) and pictured in Fig.~\ref{fig:sample}, the sample consists of a transmon qubit (red), simultaneously coupled to one detection cavity (green) and a readout cavity (purple). For both we have added a Purcell filter to protect the qubit from decay by emission into transmission lines. A weakly coupled input port (orange) allows for displacing the field in the detection cavity, a feature used for the presented Wigner tomograms.
Both the parity detection device described in this manuscript and the single photon source embedded in the on-chip switch of Ref.~\cite{Pechal2016} are mounted on the base temperature stage (20$\,$mK) of a dilution refrigerator, as shown in the wiring diagram in Fig.~\ref{fig:setup}. The switch is operated such that the radiation from the single photon source is routed to the detector for the parity measurement of sequences of single photons [Fig.~\ref{fig:1}(c)] and for the Wigner tomograms of single photon states (Fig.~\ref{fig:3}). For the cat state heralding via parity detection (Figs.~\ref{fig:4}, \ref{fig:5}), the switch routes the classical signals of the line labeled ``Coherent in'' to the detector. Both output lines are operated with a Josephson parametric amplifier as the first amplifier of the chain.

\section{Population to parity mapping}\label{app:pop-parity}
To achieve parity measurements of a signal sent towards the parity detection chip, but avoid sensitivity to other sources of qubit dephasing, such as photon shot noise, we assume that the source of qubit dephasing, leading to the observed $T_2^\star = 3.5\,\mu$s, is uncorrelated with the signal which we perform tomography on, and measure parity of in the following way.
For each set of parity measurement presented, we record two reference traces, which consist of a Ramsey sequence with pulse separation $T_w$ but no signal or displacement pulse applied. The reference for even/odd parity consists of pulses $(\pi/2,\pm\pi/2)$, respectively, applied to the transmon.  We record average qubit $|e\rangle$ state populations $P_\mathrm{e,\pm}$ in these reference traces, which differ from the ideal values of 1 and 0 by an amount $P_{T_2^*}=0.5\left[1-\mathrm{exp}(-T_w/T_2^\star)\right] \approx 12.5\,$\% as governed by the qubit coherence. The average qubit excited population measured at a given displacement $P_\mathrm{e}(\alpha)$ is in the range $(P_\mathrm{e,+},P_\mathrm{e,-})$, which we map linearly into the plotted parity value:
\begin{equation}
\langle P \rangle = \frac{P_\mathrm{e}(\alpha) - (P_\mathrm{e,+}+P_\mathrm{e,-})/2}{(P_\mathrm{e,+}-P_\mathrm{e,-})/2} \in (-1,+1),
\end{equation} an assumption that corrects for state preparation and measurement (SPAM) errors, and dephasing during the time $T_w$, under the reasonable assumption that those are independent of the signal and displacer pulses.

In a single-shot measurement, we find the correct phase in the Ramsey sequence with probability $P_\text{Ramsey}=1-P_{T_2^\star}=0.875$, and assign the correct state in the readout with probability $P_\text{ro}=(1+F_\text{ro})/2=0.97$. The correct average parity assignment is $P_\text{SSparity}=P_\text{ro}P_\text{Ramsey} +(1-P_\text{ro})(1-P_\text{Ramsey}) \approx 85\%$ for the parameters of this experiment. In the limit of low errors, that is $T_w \ll T_2^\star$, the error in the Ramsey sequence $P_{T_2^*}$ scales as $0.5 T_w/T_2^\star$, showing that average parity assignment error is reduced linearly with increasing qubit lifetime.

We measured the finite transmission efficiency $\eta=78\%$ between the source and the detection chips independently by using the nonlinear response of the single photon source and the detector as calibrated power sources~\cite{Besse2017}. An emitted sequence of $N$ single photons, where each has an independent transmission efficiency $\eta$, reaches the detector with $k$ photons with a probability given by the binomial distribution $B(k;N,\eta)={N\choose k}\eta^k (1-\eta)^{N-k}$. The expected parity, plotted in dashed red in Fig.~\ref{fig:1}(c), is given by $\langle P \rangle_\text{exp} = \sum_{k=0}^N (-1)^k B(k;N,\eta) = (1-2\eta)^N$.

\section{State reconstruction from Wigner tomograms}\label{app:wigner-model}
We use a beam-splitter model~\cite{Leonhardt1993} to account for finite transmission efficiency $\eta=78\%$ between the source and the detection chips. This models the losses by a perfect beam-splitter with transmission efficiency $\eta$, mixing the signal with a vacuum mode. The effect on the measured Wigner function is similar to that on the $Q$-function in heterodyne detection~\cite{Eichler2012}: the measured data is a convolution of the ideal Wigner function with a Gaussian kernel whose radius depends on the transmission efficiency. In mathematical terms, we measure $W'(\alpha)$ given by
\begin{equation}
W'(\alpha) = \frac{1}{\pi(1-\eta)}\int \mathrm{exp}\left(-\frac{2\eta \left| \alpha' - \alpha/\eta \right|^2}{1-\eta} \right) W(\alpha') \mathrm{d}^2\alpha'.
\end{equation}

We account for the finite overlap of the displacement pulse with the single photon waveform in the following way. We determine the waveform of the pulses emitted by our single photon source by measuring the average amplitude of $10^5$ time traces for which we prepared the photon pulse $|0{+}1\rangle$. The strong spontaneous decay to a transmission line prepares an exponential waveform with decay constant $T_p \approx 80\,$ns~\cite{Pechal2016}. We approximate the displacement pulse shape to the source waveform. A finite mode matching efficiency, given by the overlap $F_\mathrm{mm}=\sqrt{1-\epsilon^2}$ of the normalized amplitudes of the signal and the displacer pulses, leads to an effective displacement of the real signal by $\sqrt{1-\epsilon^2}\alpha$ and the displacement of an additional mode by $\epsilon\alpha$. We now assume that this additional mode is in its vacuum state, and uncorrelated with the signal. Then according to Eq.~(\ref{eq:Wigner}) the measured value of the Wigner function $W''\left(\alpha \right) = \mathrm{exp}\left(-\epsilon^2 \frac{|\alpha|^2}{2\sigma_\mathrm{vac}^2}\right) W'\left(\sqrt{1-\epsilon^2} \alpha\right)$ is separable into a product of the Wigner function of vacuum displaced by $\epsilon\alpha$ and the Wigner function of the signal displaced by $\sqrt{1-\epsilon^2}\alpha$, with $\sigma_\mathrm{vac}^2=0.5\,$photons.
We fit the value of $\epsilon$ on the measurement data taken from the single photon Fock state $|1\rangle$, see Fig.~\ref{fig:3}, and obtain a mode matching fidelity of $F_\mathrm{mm}=\sqrt{1-\epsilon^2}=84\%$. This single value for the efficiency is then used to reconstruct all most-likely density matrices. Two factors dominate the reduction of $F_\mathrm{mm}$ from its ideal value of unity. First, our data acquisition chain records data with a sampling interval of $10\,$ns, resulting in 8 data points acquired for the photon pulse shape of duration $T_p$, limiting our ability to determine the mode function with higher accuracy. Second, the input bandwidth limited by the finite coupling strength of the displacer line to the cavity, designed to be $\kappa_\text{in}=2\pi \cdot 0.5~$MHz, which is a fourth of the bandwidth of the photons, also limits the achievable mode matching bandwidth. From simulations, we expect both to reduce by a similar amount the mode matching fidelity. Pre-distortion of the pulses generated by the Arbitrary Waveform Generator (AWG), similar to used for flux pulses~\cite{Rol2019a}, faster acquisition rate for the measurement of the mode function, as well as parametrization and experimental optimization of the temporal shape of the displacement field may improve the mode matching efficiency to approach unity in future experiments.

For all single-mode Wigner tomography data sets taken, we recover the most likely density matrix of the itinerant photonic state $\rho_\mathrm{ML}$ by minimizing the norm of the difference between the corresponding Wigner function $W''(\alpha|\rho_\mathrm{ML})$ and the acquired data, enforcing a semi-positive definite density matrix with trace 1, and truncating the Hilbert space to maximally 5 photons.

\begin{figure}
\includegraphics[width=\columnwidth]{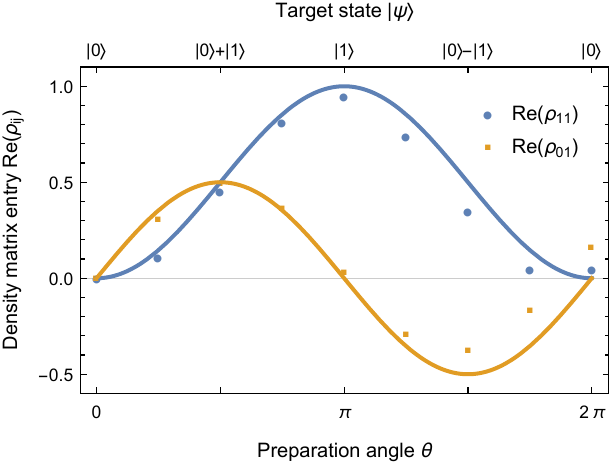}
\caption{\label{fig:rhoentries}{\bfseries Selected density matrix elements from Wigner tomography.} Real parts of the single photon population $\mathrm{Re}(\rho_{11})$ (blue) and coherence $\mathrm{Re}(\rho_{01})$ (yellow), plotted versus preparation angle $\theta$ of the source. The most likely density matrices $\rho$ are reconstructed from Wigner tomography taking losses in transmission and finite mode matching efficiency into account (see text). Solid lines represent the calculated values for the ideal states.
}
\end{figure}

As we sweep the preparation angle $\theta$ of the pulse on the source qubit, we expect to emit the state $|\gamma\rangle=\cos(\theta/2)|0\rangle + \sin(\theta/2)|1\rangle$. We plot the single photon population and the real part of the coherence in Fig.~\ref{fig:rhoentries}, together with the calculated values of $\mathrm{Re}(\rho_{01})$ and $\mathrm{Re}(\rho_{11})$ for the perfect state $|\gamma\rangle$. Here we corrected for the optical path length by applying a global phase correction minimizing the imaginary part of the coherence $\mathrm{Im}(\rho_{01})$, which has magnitudes below 0.1 in all entries (not shown).

The fidelity of the most likely density matrix $\rho_\mathrm{ML}$ with respect to the ideal density matrix $\rho_\gamma=|\gamma\rangle\langle \gamma|$ is measured by taking the trace $F=\mathrm{Tr}(\sqrt{\sqrt{\rho_\gamma}\rho_\mathrm{ML}\sqrt{\rho_\gamma}})^2$, Tab.~\ref{tab:fidelities}.
\begin{table}
\caption{\label{tab:fidelities}Fidelities $F_\theta$ for the prepared state $|\gamma\rangle=\cos(\theta/2)|0\rangle + \sin(\theta/2)|1\rangle$ (in bold, fidelities of data presented in Fig.~\ref{fig:3}).}
\begin{tabular}{r|rrrrrrrrr}
\toprule
$\theta$ & 0 & $\pi/4$ & $\pi/2$ & $3\pi/4$ & $\pi$ & $5\pi/4$ & $3\pi/2$ & $7\pi/4$ & $2\pi$  \\
$F_\theta$ & \bf{1} & 0.99 & \bf{0.98} & 0.97 &\bf{0.95} & 0.88 & \bf{0.85} & 0.88 & 0.89\\
\botrule
\end{tabular}
\end{table}
The degradation of the fidelity with increasing amplitude of the Gaussian excitation pulse is due to its rather long 180$\,$ns duration in comparison to the characteristic emission time $T_p=80\,$ns of the source. This leads to a small probability of two-photon emission~\cite{Loredo2018}. When reconstructing the state after a $\theta=2\pi$ pulse, which would ideally create the vacuum state, we find 6\% two-photon population and 4\% single photon population. This is the most likely origin of the differences between data and theory for large preparation angles, Fig.~\ref{fig:rhoentries}.

\section{Moments of input and heralded cat states}\label{app:cats-moments}
\begin{figure}[t!]
\includegraphics[width=\columnwidth]{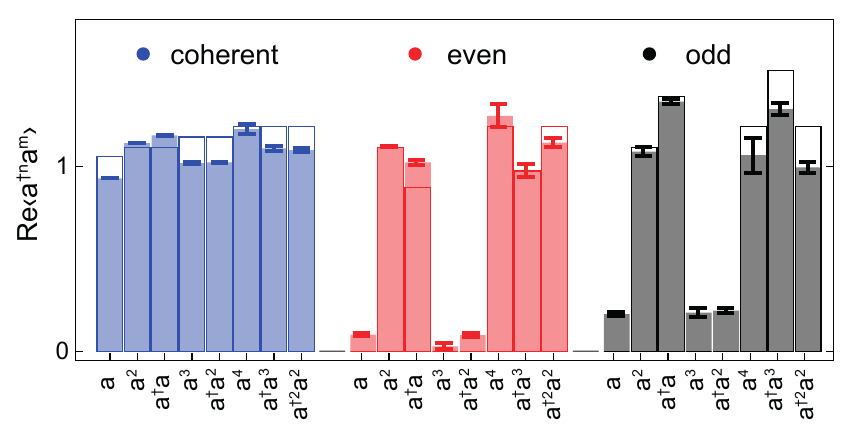}
\caption{\label{fig:homomoms}{\bfseries Optical moments of even and odd cat states.} Real part of the expectation value of the measured moments $\mathrm{Re}\langle a^{\dagger n}a^m\rangle$ (filled bars with statistical confidence interval), shown up to order $n+m\leq 4$, with $n\leq m$, for the initial state (coherent, blue) as well as conditioned on the qubit detected in the excited/ground state (even/odd cat, red/black). Wireframes represent calculated values for the ideal target states with amplitude $\alpha=1.06$. Error bars indicate the statistical standard deviation of the data.
}
\end{figure}

We measure integrated and weighted $I,Q$ quadratures of the reflected radiation using a parametric amplifier operating close to the quantum limit, with a phase-preserving gain of $18\,$dB at the cavity frequency, yielding an overall heterodyne chain detection efficiency of $\eta_\text{het} = (1+N_0)^{-1} = 23\%$, with $N_0=3.3$ photons of added noise~\cite{Eichler2012}. Together with the integrated and weighted $q$ quadrature of qubit readout, those form a three-dimensional histogram $\left\lbrace I,Q,q \right\rbrace$ from which we extract the correlations between qubit and radiation field~\citep{Eichler2012b}.
Projected on qubit readout indicating the  excited (ground) state, we evaluate the statistical moments $\langle a^{\dagger n}a^m\rangle$ of the radiation field. By applying a global phase rotation we maximize the real part of the second order moments. The relation $\mathrm{Re}\langle a^{\dagger n}a^m\rangle = \mathrm{Re}\langle a^{\dagger m}a^n\rangle$ then holds. We display the resulting measured real part of the moments up to order $n+m=4$, with $n\leq m$, averaged about 30 million times per state in Fig.~\ref{fig:homomoms}.
The imaginary parts are ideally vanishing. We observe values below 0.1 for most moments, except for the third order ones, which can reach up to 0.4. This deviation is reproduced in simulations by taking into account a small (below 5\%) deviation from the ideal $\pi$ phase shift per photon, which is consistent with the ratio of the pulse bandwidth to the cavity linewidth $\kappa_\mathrm{p}/\kappa_\mathrm{eff} \approx 7\%$, defining our precision in the acquired conditional phase over the pulse spectrum.

We observe that for the initial coherent state $|\alpha\rangle$, all moments are of order 1 (their expectation scales as $|\alpha|^{n+m}$). Once post-selected upon a given parity result, the odd order moments are heavily suppressed. This is expected as an odd number of photon annihilation/creation operators changes parity subspace. We note that the difference between the odd and even cat states are significantly larger than the statistical uncertainty, but relatively small as compared to the absolute value of the moments. The characteristic feature of an even (odd) cat state lies in the moments with $( n,m )$ both even being larger (smaller) than those with $( n,m )$ both odd.

\end{document}